\documentclass[sigconf,screen]{acmart}

\usepackage{tabularx}

\usepackage[ruled,linesnumbered]{algorithm2e}
\usepackage{url}

\AtBeginDocument{%
  \providecommand\BibTeX{{%
    \normalfont B\kern-0.5em{\scshape i\kern-0.25em b}\kern-0.8em\TeX}}}

\setcopyright{rightsretained}
\acmPrice{}
\acmDOI{10.1145/3611643.3616364}
\acmYear{2023}
\copyrightyear{2023}
\acmSubmissionID{fse23main-p1477-p}
\acmISBN{979-8-4007-0327-0/23/12}
\acmConference[ESEC/FSE '23]{Proceedings of the 31st ACM Joint European Software Engineering Conference and Symposium on the Foundations of Software Engineering}{December 3--9, 2023}{San Francisco, CA, USA}
\acmBooktitle{Proceedings of the 31st ACM Joint European Software Engineering Conference and Symposium on the Foundations of Software Engineering (ESEC/FSE '23), December 3--9, 2023, San Francisco, CA, USA}
\received{2023-02-02}
\received[accepted]{2023-07-27}

\usepackage{xparse}
\usepackage{xspace}
\usepackage{comment}
\usepackage{soul}
\usepackage{colortbl}
\usepackage{pdfpages}
\usepackage{multicol}
\usepackage{epstopdf}

\newcolumntype{P}[1]{>{\centering\arraybackslash}p{#1}}
\usepackage{makecell}
\usepackage{enumitem, kantlipsum}
\usepackage{filecontents,lipsum}
\usepackage{graphicx}
\usepackage{multirow} 
\newcommand{\tool}{\textsc{Decide}}

\usepackage{tabularray}
\UseTblrLibrary{booktabs}
\usepackage{svg}
\usepackage{booktabs}
\usepackage{array}
\definecolor{Gray}{gray}{0.9}

\definecolor{light_red}{HTML}{fddad9}

\DeclareRobustCommand{\hlred}[1]{{\sethlcolor{light_red}\hl{#1}}}

\definecolor{Gray}{gray}{0.9}

\definecolor{light_green}{HTML}{c4edc9}
\definecolor{light_red}{HTML}{fddad9}
\definecolor{light_gray}{HTML}{d5d5d5}

\DeclareRobustCommand{\hlred}[1]{{\sethlcolor{light_red}\hl{#1}}}
\DeclareRobustCommand{\hlgreen}[1]{{\sethlcolor{light_green}\hl{#1}}}
\DeclareRobustCommand{\hlgray}[1]{{\sethlcolor{light_gray}\hl{#1}}}

\begin{document}

\title{Knowledge-Based Version Incompatibility Detection \\for Deep Learning}


\author{Zhongkai Zhao}
\authornote{Work done as a remote research intern at Purdue University.}
\email{zhongkai.zhaok@gmail.com }
\affiliation{%
  \institution{Tongji University}
  \city{Shanghai}
  \country{China}
}

\author{Bonan Kou}
\email{koub@purdue.edu}
\affiliation{%
  \institution{Purdue University}
  \city{West Lafayette}
  \state{Indiana}
  \country{USA}
  \postcode{47906}
}

\author{Mohamed Yilmaz Ibrahim}
\email{ibrahi35@purdue.edu}
\affiliation{%
  \institution{Purdue University}
  \city{West Lafayette}
  \state{Indiana}
  \country{USA}
  \postcode{47906}
}

\author{Muhao Chen}
\email{muhchen@ucdavis.edu}
\affiliation{%
  \institution{University of California, Davis}
  \city{Davis}
  \state{California}
  \country{USA}
  \postcode{95616}
}

\author{Tianyi Zhang}
\email{tianyi@purdue.edu}
\affiliation{%
  \institution{Purdue University}
  \city{West Lafayette}
  \state{Indiana}
  \country{USA}
  \postcode{47906}
}

\renewcommand{\shorttitle}{Knowledge-based Version Incompatibility Detection for Deep Learning}

\begin{abstract}
    Version incompatibility issues are rampant when reusing or reproducing deep learning models and applications. Existing techniques are limited to library dependency specifications declared in PyPI. Therefore, these techniques cannot detect version issues due to undocumented version constraints or issues involving hardware drivers or OS.
To address this challenge, we propose to leverage the abundant discussions of DL version issues from Stack Overflow to facilitate version incompatibility detection. We reformulate the problem of knowledge extraction as a Question-Answering (QA) problem and use a pre-trained QA model to extract version compatibility knowledge from online discussions. The extracted knowledge is further consolidated into a weighted knowledge graph to detect potential version incompatibilities when reusing a DL project. Our evaluation results show that (1) our approach can accurately extract version knowledge with 84\% accuracy, and 
(2) our approach can accurately identify 65\% of known version issues in 10 popular DL projects with a high precision (92\%), while two state-of-the-art approaches can only detect 29\% and 6\% of these issues with 33\% and 17\% precision respectively. 
\end{abstract}




\maketitle
\section{Introduction}
Deep learning (DL) has been applied in various domains such as computer vision~\cite{furuta2019fully}, natural language processing~\cite{chen2019emoji}, and autonomous driving~\cite{li2019stereo}. Developing DL applications requires a heterogeneous DL stack, including libraries (e.g., PyTorch, TensorFlow), runtime (e.g., Python), drivers (e.g., CUDA, cuDNN), OS (e.g., Linux), and hardware (e.g., Nvidia GPU). The complex inter-dependencies between these DL stack components often result in dependency issues that are hard to diagnose and resolve~\cite{9240645}. Previous studies have shown that these issues have been identified as a major reason for build failures in DL projects, which significantly stagnates developer productivity and software reusability in DL~\cite{9240645, DBLP:journals/corr/abs-1910-11015, Chen2020ACS}.

Several techniques~\cite{horton2019dockerizeme, 10.1145/3377811.3380426, wang2021restoring,  mukherjee2021fixing, 9793962} have been proposed to detect dependency issues in Python projects, which can be applied to DL projects since most DL projects are Python-based. However, these techniques suffer from two limitations. First, all of these techniques only detect dependency issues among Python packages, with the only exception of PyEGo~\cite{9793962}, which can detect issues among Python packages, some system libraries, and Python interpreters. None of them can detect issues related to drivers, OS, and hardware. Second, these techniques rely on dependency knowledge specified in PyPI and API documentation, which has limited coverage of known version issues due to undocumented dependency and version constraints.  

Meanwhile, popular Q\&A websites such as Stack Overflow (SO) have accumulated a wealth of information about dependency issues and their solutions. 
Compared with other information sources such as PyPI, Q\&A posts are more up-to-date and comprehensive, covering various undocumented issues developers have encountered in practice. However, given the ambiguity and sophistication of natural language, extracting knowledge from free-form text is challenging. For example, a SO answer~\cite{ambiguityPaper} states that, 
{\em ``For TensorFlow 1.4, you can see only whl files up to python 3.6 are available. I am guessing that you are either using 3.7 or 3.8. That is why pip install tensorflow-gpu==1.4.0. is not working.''} To successfully extract the version knowledge, one needs to build techniques to correctly match package names with their versions and infer the (in)compatibility relationship based on the context and narrative transition across multiple sentences. 


\begin{figure*}[!t]
    \centering
    \includegraphics[width=\linewidth]{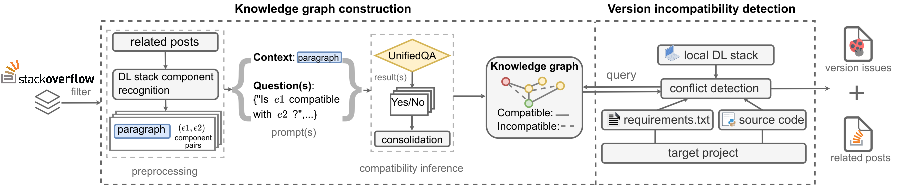}

    \caption{An overview of the knowledge graph construction and incompatibility detection process}
    \label{fig:supervised}
\end{figure*}

To address this challenge, we propose a novel approach called {\tool}, which uses a pre-trained Question-Answering (QA) model to extract version compatibility knowledge from SO posts. Figure~\ref{fig:supervised} provides an overview of {\tool}. Specifically, we reformulate the knowledge extraction task as a question-answering task: given two versioned DL stack components mentioned in a SO post, we query a QA model to predict whether they are compatible or incompatible based on the post. We use  UnifiedQA~\cite{khashabi2020unifiedqa} as the QA model, which is trained on eight large-scale datasets and has demonstrated superior performance in natural language understanding. We carefully design a set of alternative question templates based on the linguistic patterns of sentences that discuss version issues in Stack Overflow, as shown in Table~\ref{table:question_template}. By combining predictions from several alternative questions based on the loss values, {\tool} can achieve more robust predictions, which is also known as self-consistency prompting~\cite{wang2022self}.  


{\tool} further consolidates the extracted knowledge into a weighted knowledge graph, which serves as the knowledge base for version incompatibility detection. Specifically, given a DL project, \tool{} first analyzes its configuration script as well as source code to identify the required DL stack components and their version constraints. Then, it extracts DL stack information from the local machine and checks the knowledge graph for potential incompatibilities. Upon the detection of an incompatibility, \tool{} also suggests the SO posts where the knowledge is extracted from to help developers fix the issue.

We applied {\tool} to 355K SO posts that may mention version issues in deep learning, producing a large knowledge graph with 3,124 (in)compatibility relations among 48 popular DL stack components across the DL development stack. A manual analysis reveals a reasonable accuracy (84\%) of the extracted knowledge.  Furthermore, we evaluated {\tool} on 10 popular DL projects with 17 known version issues in comparison to two state-of-the-art approaches, PyEGo~\cite{9793962} and WatchMan~\cite{10.1145/3377811.3380426}. We found that {\tool} can detect 65\% of these issues with a high precision (92\%), significantly outperforming PyEGo (33\% precision and 29\% recall) and Watchman (17\% precision and 6\% recall). These results demonstrate the feasibility of knowledge extraction from Stack Overflow via a pre-trained QA model and the effectiveness of knowledge-based version incompatibility detection.

In summary, we make the following contributions:

\begin{itemize}
    \item We proposed a novel knowledge extraction paradigm that reformulates the knowledge extraction task as a question-answering task and implemented a knowledge extraction pipeline that extracts version compatibility knowledge from SO posts using a pre-trained QA model.  
    \item We developed a knowledge-based version incompatibility detection approach for DL projects.
    \item We comprehensively evaluated the quality of the constructed knowledge graph and compared \tool{} against two state-of-the-art techniques on 10 real-world DL projects.
    \item We made publicly available the first large-scale knowledge graph with 3,124 version compatibility relations, which can be used to facilitate future research on version incompatibility detection, inference, and repair for deep learning.
\end{itemize}

Our code, data, and the resulting knowledge graph have been made publicly available at \url{https://github.com/KKZ20/DECIDE}.

\section{Motivating Example}

\begin{figure}[!t]
    \centering
    \includegraphics[width=0.48\textwidth]{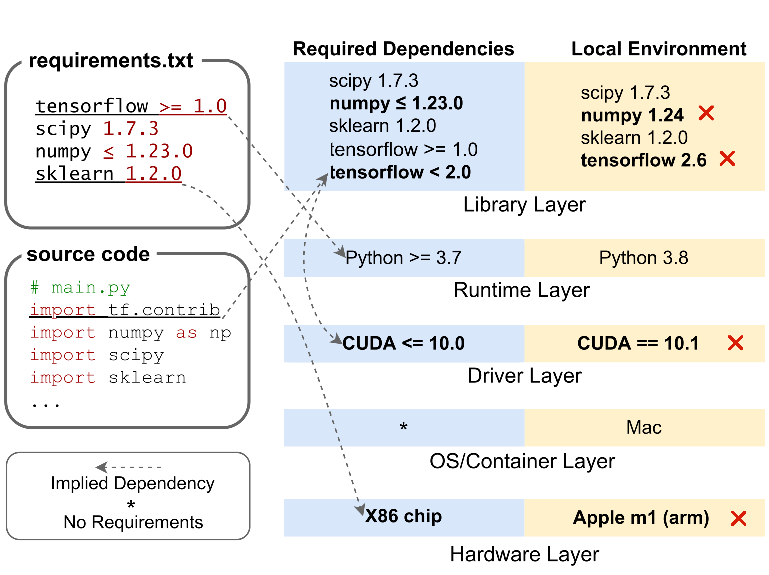}
    \caption{Required dependencies for Alice to reproduce a DL project and version incompatibilities (bold) in a local environment}
    \label{fig:motivation}
\end{figure}

Suppose Alice is a developer and she wants to reuse a TensorFlow model for sentiment analysis from GitHub.
Alice starts by installing dependencies declared in the \texttt{requirements.txt} file via 
\texttt{pip}.
Figure~\ref{fig:motivation} shows part of the \texttt{requirements.txt} file. 
She successfully installs \texttt{TensorFlow} and \texttt{NumPy}. However, when she is installing \texttt{SciPy}, \texttt{pip} throws an error message that a \texttt{numpy} version between \texttt{1.16.5} and \texttt{1.23.0} is required for \texttt{SciPy 1.7.3}, but the currently installed \texttt{numpy} version is 1.24. 
Alice solves the problem by downgrading the \texttt{NumPy} version. Yet, this is just the start of a chain of version incompatibility issues where each of them gradually reveals itself after the previous ones are resolved.

Alice continues to install \texttt{scikit-learn} with \texttt{pip}. This time, \texttt{pip} throws an error: \textit{``ERROR: Could not build wheels for NumPy, which use PEP 517 and cannot be installed directly.''} Alice finds this message confusing as she does not know what PEP 517 is. So she searches for this message online, but none of the most popular answers Google returns, such as updating \texttt{pip}, works. Alice then tries to search for the error message directly on Stack Overflow (SO). With some digging, she finds a post that suggests this error to be related to using a Macbook with an Apple M1 chip.\footnote{\href{https://stackoverflow.com/questions/70178471}{https://stackoverflow.com/questions/70178471}} 
Since Apple M1 uses the ARM architecture rather than x86, it is not compatible with the default distributions of certain libraries such as \texttt{scikit-learn}~\cite{scipy-m1}. 
According to the SO post, this issue can be solved by replacing \texttt{MacOS BLAS} with \texttt{openblas}, an open-source implementation linear algebra library that \texttt{scikit-learn} depends on and is compatible with ARM. After installing \texttt{openblas}, Alice successfully installs \texttt{scikit-learn} on Apple M1.

Now, the target DL project successfully compiles on Alice's laptop. She starts running it to train the model but immediately gets a runtime exception---``\textit{ModuleNotFoundError: No module named tf.contrib.}'' Alice is confused since she has successfully installed all library dependencies specified in \texttt{requirements.txt}. After searching online, Alice found another SO post stating that \texttt{tf.contrib} has been deprecated since \texttt{TensorFlow} \texttt{2.0} and its functionality has been migrated to the core \texttt{TensorFlow} API.\footnote{\href{https://stackoverflow.com/questions/60554127}{https://stackoverflow.com/questions/60554127}} Currently, she has \texttt{TensorFlow} \texttt{2.6} installed, which no longer has the \texttt{tf.contrib} module. Thus, she needs to downgrade her \texttt{TensorFlow} version to \texttt{1.X}. Compared with the previous incompatibility issue, this issue is much harder to diagnose, since it is not explicitly declared in the \texttt{requirement.txt} file nor clearly revealed in the error message.

Alice downgrades  \texttt{TensorFlow} to \texttt{1.15} and re-starts training. Another error occurs: ``\textit{ImportError: libcublas.so.10.0: cannot open shared object file: No such file or directory. Failed to load the native TensorFlow runtime.}'' Alice has no clue what this error means, so she searches online again. She finds yet another SO post that asks about the same error message, but it takes her a while to read and compare all 13 answer posts to this question.\footnote{\href{https://stackoverflow.com/questions/55224016}{https://stackoverflow.com/questions/55224016}} She learns that \texttt{TensorFlow} \texttt{1.15} requires \texttt{CUDA} \texttt{10.0}, a parallel computing framework to utilize NVIDIA GPUs. After manually checking the current \texttt{CUDA} version, she finds that {\texttt{CUDA} \texttt{10.2} is installed on her laptop}. She follows the instructions in an answer post to download  \texttt{CUDA} \texttt{10.0} and finally resolves the conflict. Now, with all version issues resolved, Alice can finally run the DL project to train her model. 

Instead of repeatedly searching and reading many SO posts, Alice can use \tool{} to quickly find all possible version incompatibilities between a DL project and her local machine. 
\tool{} starts by extracting version-related knowledge from thousands of SO answer posts and consolidates the extracted knowledge into a weighted knowledge graph. Then, it iteratively queries the knowledge graph to detect version incompatibilities among required DL components and locally installed components. Furthermore, to help Alice understand the root cause of the incompatibility and find solutions, \tool{} recommends a list of SO answer posts from which the incompatibility knowledge is extracted. 

Existing techniques such as WatchMan~\cite{10.1145/3377811.3380426} and PyEGo~\cite{9793962} are limited in detecting the aforementioned dependency issues for two reasons. First, most of the existing techniques only analyze dependencies among Python packages and thus cannot find issues across different DL layers. To the best of our knowledge, PyEGo~\cite{9793962} is the only technique analyzing system libraries and Python runtime in addition to Python packages. However, it still cannot analyze drivers, OS/containers, and hardware.  Second, existing techniques mainly rely on version information specified in PyPI metadata or API documentation, which is often incomplete or out-of-date. By contrast, \tool{} acquires version compatibility knowledge from abundant SO posts which are far more comprehensive and up-to-date. Therefore, it can detect a diverse set of version incompatibilities across all layers of the DL development stack (i.e., libraries, runtime, drivers, OS/containers, and hardware). Our experiment shows that \tool{} outperforms PyEGo~\cite{9793962} and WatchMan~\cite{10.1145/3377811.3380426} by at least 58.4\% in precision and 35.3\% in recall (Section~\ref{subsec: version-issue-detection}). 

\section{Problem Formulation \& Definitions}
\label{sec: formulation}
Dependency management and dependency issue detection are well-established research problems in Software Engineering (SE)~\cite{belguidoum2007dependency, esparrachiari2018tracking, cervantes2003automating, fan2020escaping, tanabe2018context}. Compared with conventional software, dependency issues in Deep Learning (DL) applications are more sophisticated due to the complex DL development stack. Thus, in this section, we first formally define the research problem and related concepts to help readers understand the scope of this project. 

Our goal is to detect potential version compatibility issues when reusing or deploying a DL project on a local machine. 
We use $ P $ to denote a DL project to be deployed and $ M $ to denote the local machine where $ P $ executes. The formal definitions are as follows: 

\vspace{1pt}
\textbf{Definition 1. (DL Stack Components):} In this work, we consider the version issues among components in the following five different DL stack layers~\cite{huang2022demystifying}: 
(1) \textit{Library Layer}: This layer contains the popular frameworks (e.g. Tensorflow, PyTorch) and other libraries (e.g. Numpy, SciPy) that a DL application directly depends on. 
(2) \textit{Runtime Layer}: This contains the execution interpreters or virtual machines of programming languages (e.g. Python interpreter, JVM).
(3) \textit{Driver Layer}: This layer includes hardware drivers and accelerated SDKs (e.g. CUDA, cuDNN)
(4) \textit{OS / Container Layer}: This includes the operating systems and other containers or virtual environments (e.g. Anaconda, Docker).
(5) \textit{Hardware Layer}: This includes the hardware and chips (e.g. CPU, GPU, TPU).

\vspace{1pt}

\textbf{Definition 2. (Local DL Stack):} We define the DL stack components $ L $ instsalled in the local machine $ M $ as $ L = \{e_1^{v_1}, e_2^{v_2}, ..., e_n^{v_n} \} $, in which $ e_i^{v_i} $ is a DL stack component $ e_i $ with version number $ v_i $.

\vspace{1pt}

\textbf{Definition 3. (Required DL Stack):} We define the DL stack components required by the DL project $ P $ as $ R = \{e_1^{c_1}, e_2^{c_2}, ..., e_n^{c_n} \} $. $ e_i^{c_i} $ is a DL stack component $e_i$ with a version constraint $c_i$, where $c_i$ is expressed in a range format $[v_{min}, v_{max}] \ (v_{min} \leq v_{max})$.

\vspace{1pt}

\textbf{Definition 4. (Version Incompatibility):} With the basic concepts defined above, we can now formulate the problem of version incompatibility.
Given a local DL stack $L$ and required components $R$ in a DL project, for any pair $(l^v, r^c)$ where $l^v \in L, r^c  \in R$, a version incompatibility issue occurs if one of the following two conditions is satisfied. First, if $l$ and $r$ refer to the same component, $v$ is not in the range of $c$. Second, if $l$ and $r$ are different, there is an implicit dependency between $r^{m}$ and $l^v$ and $m$ is not in the range of $c$.


\begin{table*}[!t]
\caption{Examples of linguistic patterns to match version-related posts}
\vspace{-6pt}
\label{table:linguistic_regex}
\resizebox{1\linewidth}{!}{
\SetTblrInner{rowsep=0pt}
\begin{tabular}{ll}
\hline
\multicolumn{1}{c}{\textsf{Regex Pattern}} & \multicolumn{1}{c}{\textsf{Matched Examples}} \\
\hline
 \makecell[l]
{\texttt{(in)?compatible*(version(s)?|(with COMPONENT\_{}NAME)?)}} 
&
\makecell[l]
{... I am using cuda 9.0 as 9.1 is not yet \hlred{compatible with tensorflow}'s \\ pre-built binary...[Post 50311325]}
\\
\hline
\makecell[l]
 {\texttt{(do not|does not|did not|don't|doesn't|didn't)?}\textbackslash s*\\ \texttt{work(s|ing|ed)?\textbackslash s*(with|for|together)}} 
 &
 \makecell[l]
{...tensorflow 1.13 \hlred{doesn't work with} cuda 10.1 because of \\ the following...[Post 55028552]} \\ 
\hline
\makecell[l]{\texttt{(be|is|are|was|were|been)(removed|deprecated|no longer support)}\\\texttt{(since|from|in))\textbackslash s}\texttt{version(s)?}}
&
 \makecell[l]
{...support for fftw was \hlred{removed in} versions of scipy >= 0.7 and \\ numpy >= 1.2... [Post 7597107]}
 \\ 
 \hline 
\makecell[l]
{\texttt{(mov|(down|up)grad)(e|ed|ing)\textbackslash{}s*(your)?\textbackslash{}s(COMPONENT\_{}NAME)}\\ \textbackslash{}s*\texttt{(version)?\textbackslash{}s*(from|to)?}}
&
\makecell[l]
{...\hlred{downgrade numpy version from} 1.17.2 to 1.16.4 will resolve \\ issue with tensorflow...[Post 61817557]} \\ 
\hline
\makecell[l]{\texttt{(latest|new|earlier|older|previous|later|recent}\\\texttt{|minimum|maximum)\textbackslash s}\texttt{(version(s)?)\textbackslash s(of)?}}
&
 \makecell[l]
{...what changed: the \hlred{latest version of} numpy requires python 3.5+, \\ hence the error message... [Post 57734033]}
 \\ 
 \hline
 
\end{tabular}}
\end{table*}

\vspace{1pt}

While the first condition is easy to check, the second condition is sophisticated and challenging, since it requires identifying implicit dependencies between DL stack components. Existing approaches rely on program analysis or dependency graphs from PyPI or the Python official website to identify implicit dependencies. However, these approaches cannot handle implicit dependencies across DL stack layers or capture undocumented dependencies. In this work, we propose to utilize the rich information shared on Stack Overflow, which captures various and up-to-date version incompatibility issues in the real world. We extract and represent such information in a weighted knowledge graph, which is defined below.

\vspace{1pt}

\textbf{Definition 5 (Weighted Knowledge Graph): } A weighted knowledge graph is defined as 
$ KG = <N, E> $ where 
$ N = \{ n_i^v \ | \ i \ge 0 \} $
is a set of nodes denoting the DL stack components with their version numbers, and 
$ E = \{ e_j^w \ | \ j \ge 0 \} $
is a set of edges representing the compatible or incompatible relationship between two DL stack components. Each edge is labeled with a normalized weight $ w $, which captures the confidence of this knowledge (detailed in Section~\ref{sec:consolidate}). 



\section{Knowledge Graph Construction}
This section presents how {\tool} extracts version compatibility knowledge from SO posts to build a knowledge graph.

\subsection{Data Collection and Filtering}
\label{sec:data}
We downloaded the Stack Exchange Data Dump~\cite{SOdump} with 53 million Stack Overflow posts from July 31, 2008 to September 5, 2021. Since our purpose was to extract version compatibility knowledge related to deep learning, we first filtered the SO data dump to find relevant SO posts. To do this, we manually identified a set of SO tags related to deep learning. 
Specifically, the first author manually went through 798 popular SO tags (i.e., tags with more than 10K questions) from all the 63,715 tags in Stack Overflow and created an initial lexicon with 12 keywords related to deep learning. To get a more accurate tag list, the first author then searched the remaining 798 SO tags and found 525 tags containing at least one of these 12 keywords. 
Then, he inspected all these tags and selected a final set of 46 SO tags related to deep learning.\footnote{The complete list of tags can be found at \url{https://github.com/KKZ20/DECIDE/blob/main/DECIDE/docs/SO_tags.json}} We filtered the SO posts to only retain posts tagged with at least one of the 46 tags. 4.9M posts remained after this step.

Then, we performed another round of filtering to find SO posts that may mention version compatibility issues. From the DL-related posts obtained from the previous step, the first author searched \textit{``version incompatibility''} and manually inspected the first 150 posts from the search results. Then, he summarized 22 linguistic patterns from the sentences mentioned version issues in these 150 posts. Table~\ref{table:linguistic_regex} shows five examples of the 22 linguistic patterns.\footnote{The complete list of linguistic patterns can be found at  \url{https://github.com/KKZ20/DECIDE/blob/main/DECIDE/docs/Linguistic_patterns.txt}} We further filtered the DL-related SO posts with these patterns to find SO posts that mention version issues. After this step, 549K posts were retained.

Furthermore, to ensure the quality of the SO posts, we only kept posts that were marked as accepted answers, as well as posts whose vote score (i.e., upvotes minus downvotes) was above one. This resulted in 355K posts, which formed the information source to extract version compatibility knowledge.

To evaluate the accuracy of our data collection process, we randomly sampled 384 posts from the filter results. This sample size is statistically meaningful with a 95\% confidence interval.
The first two authors independently inspected these posts and found 326 of them indeed contained version compatibility information. The Cohen's Kappa score of this evaluation is 0.85. In other words, our data collection pipeline identified SO posts with version compatibility knowledge with 84.9\% accuracy, which is reasonable for knowledge extraction.

\subsection{DL Stack Component Recognition}
\label{sec:DL Stack Component Recognition}

During the manual inspection in the previous step, we observed that not all paragraphs in a version-related SO post mentioned version compatibility information. To improve the knowledge extraction efficiency, we designed a filtering mechanism in {\tool} to locate paragraphs that may mention version compatibility information. The key insight is that version incompatibility involves two DL stack components and their versions. So {\tool} only selected paragraphs that mention at least two different versioned components.

In the current implementation, {\tool} supports the recognition of 48 popular components across the five layers in a DL stack. These components were manually identified by the first author from the 200 posts with the highest score (i.e., upvotes minus downvotes) among all posts obtained from the previous step. In addition, the first author also added synonyms or aliases for these components to improve the accuracy of DL stack component recognition. 
Table~\ref{table:stackcomponent} shows some of the 48 components and their synonyms.
\footnote{A complete list of 48 DL stack components can be found at  \url{https://github.com/KKZ20/DECIDE/blob/main/DECIDE/docs/DL_Stack_Components.txt}} Currently, we only consider these popular components. One can easily extend them by adding more components, either manually or from an existing lexicon. Adding more components does not induce any additional effort to adapt the following steps, since the following steps are designed as a general process for any DL stack components.

\begin{table}[!h]
\caption{Examples of the recognized DL stack components (names after | are aliases)}
\label{table:stackcomponent}
\resizebox{\linewidth}{!}{
\SetTblrInner{rowsep=0pt}
\begin{tblr}{
    colspec={Q[l]|l},
}
\hline
\SetCell[c=1]{l}\textsf{Layer} & \SetCell[c=1]{l}\textsf{DL Stack Components} \\
\hline
\SetCell[c=1]{l}Library &
\SetCell[c=1]{l}Tenforflow \& tf, Numpy \& np, PyTorch, scikit-learn \& sklearn \\
\hline
\SetCell[c=1]{l}Runtime &
\SetCell[c=1]{l}Python \\
\hline
\SetCell[c=1]{l}Driver &
\SetCell[c=1]{l}CUDA, cuDNN \\
\hline
\SetCell[c=1]{l}OS/Container &
\SetCell[c=1]{l}Ubuntu, Windows, MacOS, Debian, Anaconda \\
\hline
\SetCell[c=1]{l}Hardware &
\SetCell[c=1]{l}Apple M1, ARM, AMD \\
\hline
\end{tblr}
}
\end{table}

Based on the 48 DL stack components, {\tool} performs keyword matching to identify paragraphs that mention at least two different components. Specifically, given a SO post, {\tool} first removes code snippets wrapped in \texttt{<pre>} tags from the post to focus on the natural language text data. It keeps the inline code elements wrapped in \texttt{<code>} tags to preserve the text flow. Then, {\tool} splits the preprocessed text into paragraphs by line breaks. When matching DL stack components, {\tool} performs case-insensitive keyword matching. Furthermore, we designed three regex patterns to identify version numbers mentioned in a paragraph. Table~\ref{table:version regex} shows the three regex patterns and some examples of matched version numbers. 

\begin{table}[!h]
\caption{Version matching patterns}
\vspace{-6pt}
\label{table:version regex}
\resizebox{\linewidth}{!}{
\SetTblrInner{rowsep=0pt}
\begin{tblr}{
    colspec={Q[l]|l},
}
\hline
\SetCell[c=1]{c}\textsf{Regex Pattern} & \SetCell[c=1]{c}\textsf{Matched Examples} \\
\hline
\SetCell[c=1]{l}\texttt{v\{0,1\}\textbackslash{}d+(\textbackslash.d+)\{1,2\}} & 
\SetCell[c=1]{l}\texttt{3.7, 2.4.3, v2.3, v1.13.5} \\ 
\SetCell[c=1]{l}\texttt{v\{0,1\}\textbackslash d+(\textbackslash.\textbackslash d+)\{0,1\}(\textbackslash.x)\{0,1\}} &
\SetCell[c=1]{l}\texttt{3.x, 1.3.x, v1.x, v2.2.x} \\ 
\SetCell[c=1]{l}\texttt{(COMPONENT)(-| |\_{})}\texttt{v\{0,1\}\textbackslash d+} &
\SetCell[c=1]{l}\texttt{python v3, cuda-8, Windows 64} \\ 
\hline
\end{tblr}
}
\end{table}

\begin{figure}[t]
    \centering
    \includegraphics[width=0.45\textwidth]{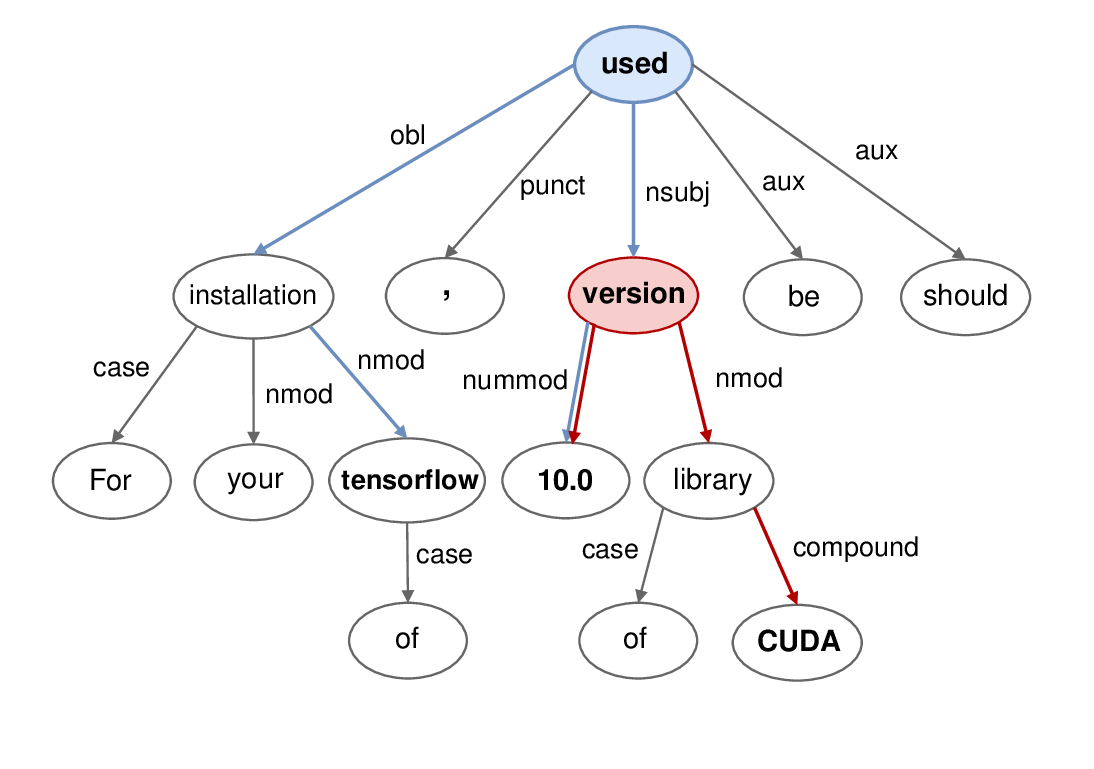}
    \vspace{-23pt}
\caption[Dependency tree where the depth of the lowest common ancestor for 10.0 and CUDA is 1 while the depth for the lowest common ancestor for 10.0 and tensorflow is 0.]{Dependency tree where the depth of the lowest common ancestor for 10.0 and CUDA is 1 while the depth for the lowest common ancestor for 10.0 and tensorflow is 0.\footnotemark}
    \label{fig:DependencyTree}
\end{figure}

After extracting a set of DL stack components and version numbers in a paragraph, \tool{} makes the best effort to match a component with its version. 
We formulate this matching problem as a weighted stable matching problem~\cite{wikipedia-contributors-2023}. The stable matching algorithm tries to establish a one-on-one matching between each component and each version while maximizing the depth of
the lowest common ancestor of every matched component and version in the dependency tree of a sentence. This measurement is more robust than the token-level distance, since we observe that the closest version number to a component in a sentence might refer to another component given the complex grammar structure in natural language. Consider this example: \textit{``For your installation of tensorflow, 10.0 version of CUDA library should be used''}. Though \texttt{10.0} is closer to \texttt{tensorflow}, it is actually the version of \texttt{CUDA}. Figure~\ref{fig:DependencyTree} shows the dependency tree of this sentence. The lowest common ancestor of  \texttt{10.0} and \texttt{tensorflow} is \texttt{used} (depth 0), while the lowest common ancestor of \texttt{10.0} and \texttt{CUDA} is \texttt{version} (depth 1). Thus, \tool{} matches\texttt{10.0} with \texttt{CUDA}. Specifically, {\tool} uses  Stanza~\cite{qi2020stanza} to perform dependency parsing.

After the matching process, if a component in the library, runtime, driver, or OS/Container layer does not have a matched version, it will not be considered for compatibility inference in the next stage. Yet if a component in the hardware layer does not have a matched version, it will still be considered for compatibility inference in the next stage. This design choice is based on our observation that developers do not always report hardware versions when mentioning version issues on Stack Overflow. Consider the example: \textit{``I face the same problem on macOS 11.6.8 (BigSur) with the Apple M1 chip''}.\footnote{\href{https://stackoverflow.com/questions/73605384}{https://stackoverflow.com/questions/73605384}} In this sentence, macOS will be matched with 11.6.8, while Apple M1 is a component with no version number. In the end, {\tool} identifies a total of 1,700 paragraphs from 2,018 SO posts that mention at least two different versioned components.

Overall, our matching algorithm makes the best effort to match a DL stack component and its version. Given the ambiguity and complexity of natural language, we cannot always guarantee the correctness of matching. Yet, with the careful design above, we argue that our matching algorithm works in most cases. To confirm this, we randomly sampled 384 SO posts from the 355K posts from the previous step to evaluate our matching algorithm. 
This sample size is statistically meaningful with a 95\% confidence interval. The first two authors independently checked the matching results from each post to determine whether DL stack components were correctly matched with their versions. Then, they compared their analysis results and resolved any disagreements. The Cohen’s Kappa score was 0.86. Overall, our matching algorithm achieved a high accuracy of 87.4\%. This result indicated that our matching algorithm worked fine for most cases.

\subsection{Compatibility Inference via a Pre-trained QA Model}
After recognizing two DL stack components and their versions in a paragraph, {\tool} infers the compatibility relationships between them based on the information in the paragraph. Analyzing free-form text in SO posts is challenging due to the ambiguity in their narratives as well as their sophisticated structures. Prior work that uses linguistic patterns or rules to extract programming knowledge falls short in reasoning the deep semantics in natural language. In this work, we propose to reformulate this compatibility relationship classification task as a Question-Answering (QA) task and then use a pre-trained QA model to solve the task. Specifically, \tool{} uses UnifiedQA~\cite{khashabi2020unifiedqa} as the pre-trained QA model. UnifiedQA is a large model with 3 billion parameters and it is pre-trained on eight datasets. It has been demonstrated to understand deep semantics in natural language and achieve state-of-the-art performance in multiple QA benchmarks~\cite{khashabi2020unifiedqa}. UnifiedQA takes two inputs---a {\em question} and a {\em context document} from which the answer is extracted. \tool{} uses the paragraph where two versioned components are recognized as the context document and then asks UnifiedQA a yes-or-no question to infer the compatibility relationship between the two components. Figure~\ref{table:qa_example} shows two QA examples from two real SO posts---\href{https://stackoverflow.com/questions/60526751}{\hlgray{[Post 60526751]}} and \href{https://stackoverflow.com/questions/55028552}{\hlgray{[Post 55028552]}}. 

\footnotetext{The POS tags are omitted in this dependency tree for presentation. The definition of each edge label can be found at \href{https://universaldependencies.org/u/dep/all.html}{https://universaldependencies.org/u/dep/all.html}}

\begin{table}[]
\captionsetup{skip=5pt}
\begin{center}
\begin{tabularx}{0.45\textwidth}{X}
\hline
\multicolumn{1}{|X|}{\textbf{Context:} \textit{import tensorflow} issue has been resolved by changing python from 32 bit to 64 bit and python version must be 3.5-3.7 because 3.8 is not compatible for installing tensorflow through:
\textit{pip install tensorflow==1.5.0}.

\textbf{Question:} 
\hlgreen{Is \texttt{python} \texttt{3.7} compatible with \texttt{tensorflow} \texttt{1.5.0}?}
}  \\\hline
\end{tabularx}
\vspace{0.2em}
\begin{tabularx}{0.45\textwidth}{X}
\hline
\multicolumn{1}{|X|}{\textbf{Answer from UnifiedQA:} Yes.}\\ \hline
\end{tabularx}

\begin{tabularx}{0.45\textwidth}{X}
\hline
\multicolumn{1}{|X|}{\textbf{Context:} \texttt{tensorflow} \texttt{1.13} doesn't work with \texttt{cuda 10.1} because of the following: \textit{``ImportError: libcublas.so.10.0: cannot open shared object file: No such file or directory''}. \texttt{tensorflow} is looking for \texttt{libcublas.so.10.0} whereas \texttt{cuda} provides \texttt{libcublas.so.10.1.0.105}.

\textbf{Question:} 
\hlgreen{Does \texttt{tensorflow} \texttt{1.13} work with \texttt{cuda} \texttt{10.1}?}
}  \\\hline
\end{tabularx}
\vspace{0.2em}
\begin{tabularx}{0.45\textwidth}{X}
\hline
\multicolumn{1}{|X|}{\textbf{Answer from UnifiedQA:} No.}\\ \hline
\end{tabularx}

\end{center}
\captionof{figure}{QA examples for version compatibility inference}
\label{table:qa_example}
\end{table}

The selection of appropriate question prompts for QA models has a noticeable impact on model performance~\cite{radford2021learning}. 
Hence, we carefully crafted eight question templates  (Table~\ref{table:question_template}) based on the 22 linguistic patterns identified in the post filtering procedure (Table~\ref{table:linguistic_regex}). We experimented with all eight question templates and some of their combinations on UnifiedQA. The experiment results indicate the best performance of UnifiedQA is achieved by combining Q1 and Q2. More details on evaluations of different question templates can be read in Section~\ref{sec: QTanalysis}.

\begin{table}[]
\caption{Question templates for version knowledge extraction}
\begin{center}
\begin{tabular}{{c|l}} 
 \hline
\cellcolor[gray]{0.7}&Q1: Is $(e_A, v_A)$ compatible with $(e_B, v_B)$? \\ 
\cellcolor[gray]{0.7}&Q2: Is $(e_A, v_A)$ not compatible with $(e_B, v_B)$? \\ 
\cellcolor[gray]{0.7}&Q3: Does $(e_A, v_A)$ support $(e_B, v_B)$? \\
\cellcolor[gray]{0.7}&Q4: Does $(e_A, v_A)$ not support $(e_B, v_B)$? \\
\cellcolor[gray]{0.7}&Q5: Does $(e_A, v_A)$ require $(e_B, v_B)$? \\
\cellcolor[gray]{0.7}&Q6: Does $(e_A, v_A)$ not require $(e_B, v_B)$? \\
\cellcolor[gray]{0.7}&Q7: Does $(e_A, v_A)$ work with $(e_B, v_B)$? \\
\multirow{-8}{*}{\cellcolor[gray]{0.7}\rotatebox{90}{\textbf{Question Templates}}}&Q8: Does $(e_A, v_A)$ not work with $(e_B, v_B)$? \\\hline
\end{tabular}
\label{table:question_template}
\end{center}
\end{table}

\subsection{Knowledge Consolidation}
\label{sec:consolidate}

The exacted version compatibility relationships may contain redundancies and inconsistencies, as the relevant information can be presented in multiple posts. To eliminate redundancies and reconcile conflicts, \tool{} consolidates relations between the same pairs of versioned DL stack components by calculating the confidence weight for each relation. For a pair of versioned components, let $ \#Compatible $ denote the number of posts that {\tool} infers a compatibility relationship between them, and $ \#Incompatible $ denote the number of posts that {\tool} infers an incompatible relationship between them. We define the confidence weight of the relationship between two versioned components as follows: 
\begin{equation}
    conf = \frac{\#Compatible -\#Incompatible}{\#Compatible + \#Incompatible}
\end{equation} 

If $conf$ is a positive number, it implies a compatible relationship. Otherwise, it implies an incompatible relationship. Relationships with a neutral confidence weight ($ conf = 0 $) are discarded.

\begin{figure}[!t]
    \centering
    \includegraphics[width=0.40\textwidth]{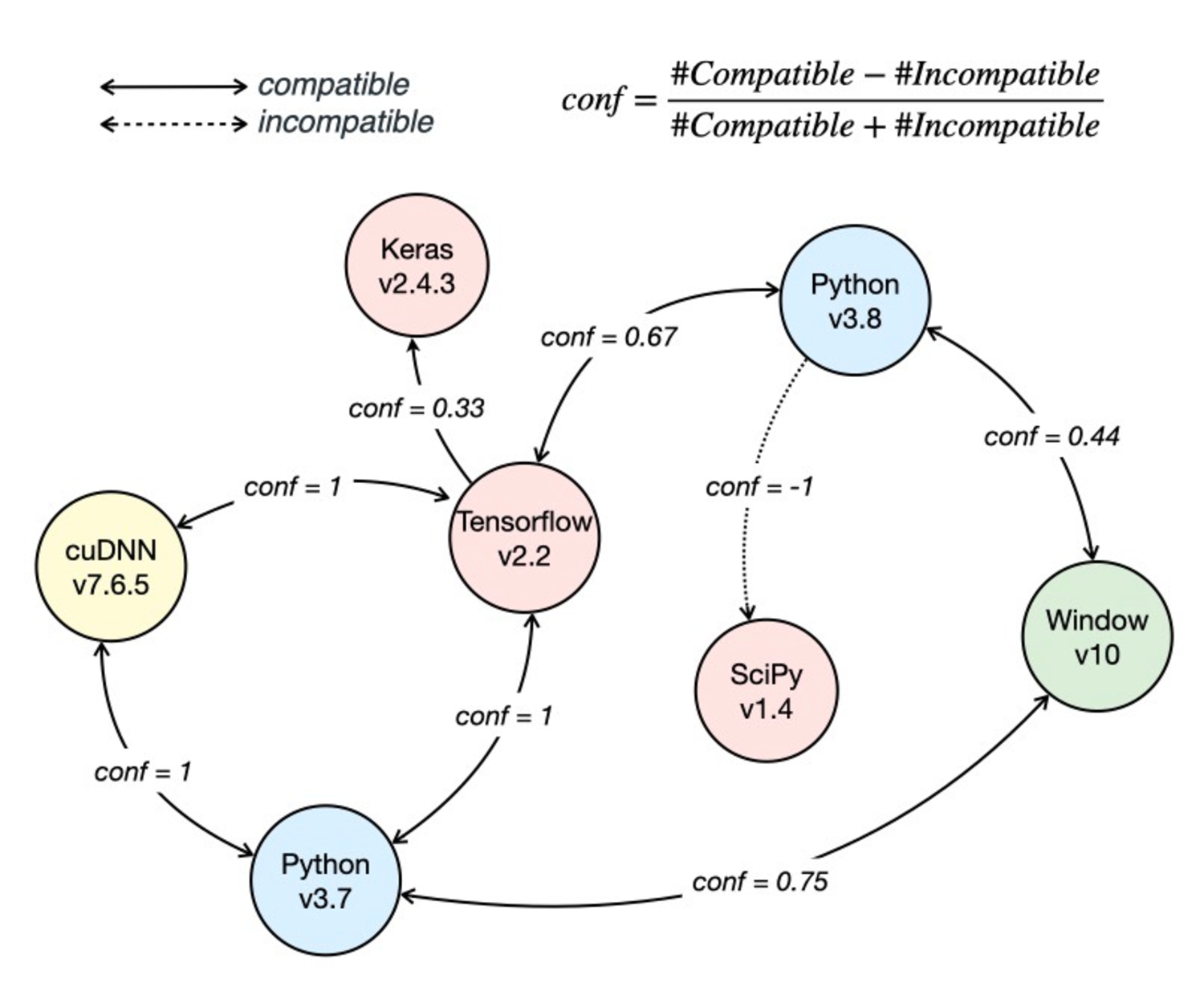}
    \caption{Part of the resulting knowledge graph}
    \label{fig: Knowledge Graph}
\end{figure}

 After the knowledge consolidation process, \tool{} produces a knowledge graph consisting of 1,431 nodes and 3,124 edges. Each node in the graph represents a unique versioned DL stack component, while each edge represents an (in)compatibility relationship between two components with a confidence weight. An illustration of the knowledge graph is provided in Figure~\ref{fig: Knowledge Graph}, where components of different layers in the DL development stack are denoted by different colors. For example, \texttt{Keras} \texttt{2.4.3}, \texttt{Tensorflow} \texttt{2.2}, and \texttt{SciPy} \texttt{1.4} are colored red in Figure~\ref{fig: Knowledge Graph}, since they come from the library layer. All the relations in the knowledge graph are weighted. In Figure~\ref{fig: Knowledge Graph}, the relation between \texttt{Python} \texttt{3.8} and \texttt{Tensorflow} \texttt{2.2} has a confidence weight of $ 0.67 $, indicating a compatible relationship. This weight is calculated based on Equation 1, where 10 SO posts mention these two library versions are compatible ($\#Compatible$) and 2 posts mention the opposite  ($\#Incompatible$). This pipeline of building the weighted knowledge graph can be adapted to extract open-ended knowledge for different entities from future SO posts.

\section{Compatibility Issue Detection}
Given a DL project to reuse and a local environment, \tool{} performs version incompatibility detection based on the knowledge graph in three steps: (1) identifying the required DL stack from the given DL project, (2) identifying the local DL stack from the local environment, and (3) querying the knowledge graph to detect potential version incompatibility if installing the required components on the local environment. Each step is described below. 

\subsection{Required DL Stack Identification}
\tool{} first identifies required DL stack components from the configuration file and source code. As a common practice, developers usually specify required dependencies for a project in a configuration file named \texttt{requirements.txt}~\cite{abate2011predicting, 10.1145/3377811.3380426}. \tool{} collects the names and version constraints of the packages and libraries specified in \texttt{requirements.txt}. However, as the components listed in \texttt{requirements.txt} can be incomplete, \tool{} further performs static analysis to obtain the full list of Python packages used by the project. Specifically, \tool{} parses all the Python files in the project into Abstract Syntax Trees (ASTs)~\cite{ast-python} and extracts package names specified in the \textit{import} statements.

\subsection{Local DL Stack Identification}
Following Definition 1 in Section~\ref{sec: formulation}, \tool{} collects the version information of the components in five DL stack layers. Note that a developer may deploy the given DL project in either a native Python environment or a virtual environment managed by Anaconda~\cite{conda}. \tool{} will first detect which environment is used by running \texttt{``echo \$CONDA\_PREFIX''}. If this command outputs a non-empty path of a conda environment, then it is a virtual environment. Otherwise, it is a native environment. Depending on the local environment, {\tool} will invoke different system commands to gather the version information of the local DL stack. 

We describe how {\tool} gathers the version information in each layer below:
\begin{enumerate}[leftmargin=*]
    \item \emph{Library}:  If the local environment is a native Python environment, 
    \tool{} uses \texttt{pip}~\cite{pypi}, the default Python package manager, to gather locally installed Python packages and their versions via \texttt{``pip freeze''}. Then, \tool{} gathers the version information of system libraries (e.g., GCC) by directly calling corresponding system commands (e.g., \texttt{``gcc -{}-version''})  and parsing the outputs. In the current implementation, {\tool} only extracts the version information of the popular system libraries described in Section~\ref{sec:DL Stack Component Recognition}. If the local environment is a virtual  environment managed by  Anaconda~\cite{conda}, {\tool} will use the \texttt{conda} command to collect version information---\texttt{``conda list''}. While \texttt{pip} is centered around Python packages, \texttt{conda} supports the management of system libraries (e.g., \texttt{gcc}) and hardware drivers (e.g., \texttt{cuDNN}) in addition to Python packages. Thus, {\tool} can use \texttt{``conda list''} to gather the version information of both Python packages and system libraries in a single command execution. 
    
    \item \emph{Runtime}: Since we focus on Python-based DL projects in this paper, \tool{} gathers the runtime Python interpreter version by invoking \texttt{``python -{}-version''} and parsing the output. This command works in both  native and virtual environments. 

    \item \emph{Driver}: If the local environment is native, \tool{} will collect the version information of hardware drivers and toolkits by directly invoking the corresponding command for each driver component described in Section~\ref{sec:DL Stack Component Recognition}, e.g., \texttt{``nvcc -v''} for runtime CUDA versions, \texttt{``nvidia-smi''} for Nvidia GPU driver versions. If the local environment is virtual, \tool{} will use \texttt{``conda list''} to gather the version information of hardware drivers and toolkits.
    \item \emph{OS / Container}: \tool{} uses APIs in Python \texttt{os} module or the command of each OS/Container component to gather their version information, e.g., python \texttt{os} module for OS version, \texttt{``conda -{}-version''} for Anaconda version. 
    \item \emph{Hardware}: \tool{} gathers hardware information by parsing the outputs of system commands, e.g., \texttt{``uname -a''} for the CPU architecture, \texttt{``nvidia-smi''} for the Nvidia GPU model.
\end{enumerate}

In the current implementation, we focus on 48 popular DL stack components in the five layers. These components are described in Section~\ref{sec:DL Stack Component Recognition} and a complete list can be found in the Supplementary Material. Our approach can be easily extended to collect version information for other components by adding the corresponding system commands as described above.


\subsection{Version Incompatibility Detection}

After collecting version information of required DL stack components and local DL stack components, \tool{} detects version incompatibilities in the following steps. For each required DL stack component $ r_i^{c_i} $, \tool{} first checks if it is installed in the local machine. 
If $ r_i $ is locally installed with a version $ v_i $ and $ v_i $ is in the range $ c_i $, \tool{} considers there is no version issue and moves on to the next component. Otherwise, \tool{} reports a dependency issue and infers the correct version of $ r_i $. To infer the correct version, \tool{} first queries the knowledge graph $ KG $ to get a set of candidate versions $ S_i = \{ s_{1}, s_{2}, ... s_{n} \} $ sorted in the ascending order. For each candidate in $ S_i $, {\tool} check if it is compatible with each installed component in the local DL stack by querying the knowledge graph. 
Then, \tool{} chooses the candidate with the latest version,  $s_m$ (m<=n), which is compatible with every other component in the local stack. It then pushes $ r_i^{s_{m}} $ into a stack $ V $ to keep track of the installed or fixed component versions. \tool{} repeats the previous steps to check all the required DL stack components. During this process, if {\tool} cannot infer a component version that satisfies the required version constraint and is compatible with all other components in the local stack, {\tool} first backtracks to the previous inference step and pops up the inferred component in $ V $. It will then pick another version candidate of that component in $S_i$ and continue the process. If $V$ is empty, then {\tool} reports no solution can be found. During this process, every time {\tool} reports a version issue or recommends a compatible version, it also reports the SO posts where the (in)compatibility knowledge is extracted to help developers understand the issue or recommendation. For each project in the evaluate benchmark (detailed in Section~\ref{subsec: version-issue-detection}), \tool{} is able to detect and report version issues within 1 minute without using GPU.

\section{Evaluation}
We conduct experiments to answer five research questions below:

\begin{itemize}[leftmargin=*]
\item \textbf{RQ1}: How effectively can \tool{} detect version compatibility issues in real DL projects?
\item \textbf{RQ2}: How accurate is the extracted knowledge in the resulting knowledge graph produced by {\tool}?
\item \textbf{RQ3}: How accurately can the pre-trained QA model in \tool{} infer compatibility relations between versioned DL components? 
\item \textbf{RQ4}: To what extent can different question templates affect the accuracy of the pre-trained QA model in \tool{}?
\item \textbf{RQ5}: To what extent can different knowledge consolidation strategies affect the accuracy of the resulting knowledge graph?
\end{itemize}


\subsection{Version Incompatibility Detection}
\label{subsec: version-issue-detection}

\subsubsection{Benchmark Construction} To evaluate the performance of \tool{} on real-life deep learning projects, we created a benchmark consisting of 10 popular DL projects from GitHub. To create this benchmark, we first searched for deep learning projects on GitHub that have at least 100 stars and have a \texttt{requirements.txt} file. Given the search results, we randomly sampled one deep learning project at a time and manually checked whether it contains the implementation of a deep learning model. Then, we manually reproduced it on our local machine to check if it indeed contains version issues. Our local machine includes a Ubuntu 18.04 LTS with an Intel x86-64 CPU and one RTX A5000 GPU. The default Python version is 3.6 and the native CUDA version is 11.2.  We continued this process until we found 10 projects with at least one version issue on our local machine. 

Table~\ref{table: benchmark statistics} shows the names and statistics of these 10 DL projects. On average, these projects have 1,647 stars, 1,321 lines of code, and 1.7 version issues. Seven of them are computer vision models, two are natural language processing models, and one is a graphical neural network. 
To ensure a comprehensive identification of version incompatibility issues, we reproduced each project, ensured that the model could run successfully without errors, and documented all encountered version issues during the reproduction process.
We get 17 version issues from the 10 projects in total. 
4 of these issues involve components at the same level, while 13 issues involve components between two different layers, e.g., incompatibility between TensorFlow and CUDA. Specifically, 17 issues involve a library component, 5 involve a runtime component, 10 involve a driver component, 1 involves an OS/Container component, and 1 involves a hardware component. Details of these version issues can be found in our artifact~\cite{artifact}.

\begin{table}[!h]
\centering
\caption{Benchmark project statistics}
\label{table:benchmark_statistics}
\resizebox{\linewidth}{!}{
\begin{tblr}{
    row{1} = {font=\bfseries},
    row{2-Z} = {rowsep=.3pt},
}
\toprule
\SetCell[c=1]{c} \textsf{Project Name} & \textsf{Star} & \textsf{LOC} & \textsf{Domain} & \textsf{\# Issue} \\
\midrule
\href{https://github.com/tkipf/gcn}{tkipf/gcn} & 6,785 & 707 & GNN & 2 \\
\href{https://github.com/chonyy/AI-basketball-analysis}{chonyy/AI-basketball-analysis} & 853 & 648 & CV & 2 \\
\href{https://github.com/fidler-lab/polyrnn-pp}{fidler-lab/polyrnn-pp} & 732 & 723 & CV & 2 \\
\href{https://github.com/taki0112/StyleGAN-Tensorflow}{taki0112/StyleGAN-Tensorflow} & 212 & 1,625 & CV & 1 \\
\href{https://github.com/rishizek/tensorflow-deeplab-v3-plus}{rishizek/tensorflow-deeplab-v3-plus} & 825 & 1,632 & CV & 1 \\
\href{https://github.com/cfernandezlab/CFL}{cfernandezlab/CFL} & 101 & 1,457 & CV & 1 \\
\href{https://github.com/NVlabs/noise2noise}{NVlabs/noise2noise} & 1,297 & 3,273 & CV & 1 \\
\href{https://github.com/localminimum/QANet}{localminimum/QANet} & 990 & 1,805 & NLP & 1 \\
\href{https://github.com/kaonashi-tyc/Rewrite}{kaonashi-tyc/Rewrite} & 735 & 544 & CV & 5 \\
\href{https://github.com/gaussic/text-classification-cnn-rnn}{gaussic/text-classification-cnn-rnn} & 3,943 & 802 & NLP & 1 \\
\midrule
AVG & 1,647 & 1,321 & / & 1.7 \\
Median & 839 & 1,129 & / & 1 \\
\bottomrule
\end{tblr}
}
\end{table}


{\subsubsection{Comparison Baselines} We compare our tool against two state-of-the-art approaches, PyEGo~\cite{9793962} and Watchman~\cite{10.1145/3377811.3380426}:}
{\begin{itemize}[leftmargin=*]
    \item \textbf{\textit{Watchman}}~\cite{10.1145/3377811.3380426} extracts dependency relations between third-party Python packages from PyPI documentation and uses the extracted dependency relations to detect dependency conflicts of installed packages in a Python project. We used the public web interface of WatchMan from its official website~\cite{watchman-interface} for evaluation. This web interface takes a text file of installed packages as input and outputs potential dependency issues among the packages. To couple with this input format, for each DL project in the benchmark, we listed all locally installed packages and their versions, as well as any missing package and its version constraint specified in \texttt{requirements.txt}, in a text file and uploaded it to the web interface of Watchman. The version issues reported by Watchman will be used to compare with the ground truth.
    \item \textbf{\textit{PyEGo}}~\cite{9793962} extracts dependency relations between Python packages, Python interpreters, and system libraries based on PyPI documentation, Python documentation, etc.
    Given a Python project, it uses the extracted knowledge to infer the latest compatible versions of required dependencies for the project based on the source code. We used the PyEGo implementation from its GitHub repository~\cite{pyego-github} for evaluation. Given a DL project, we used PyEGo to infer a set of compatible versions for the required DL components. Then, we compared the inferred versions of required components with the versions of locally installed components to detect version issues.
\end{itemize}}



\subsubsection{Evaluation Results} 
\begin{figure}[!t]
    \centering
    \includegraphics[width=0.45\textwidth]{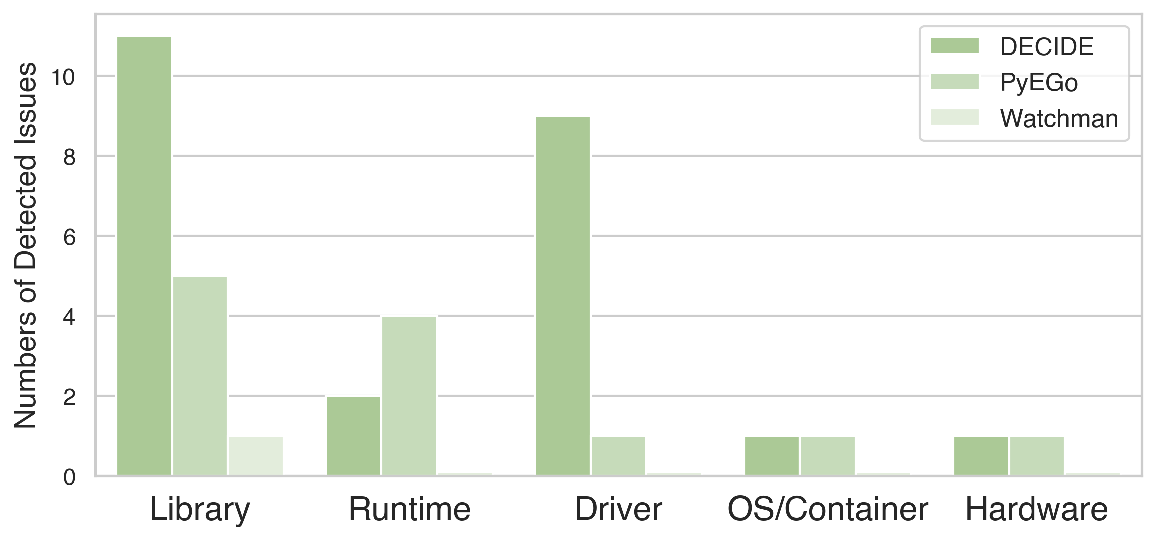}
    \caption{The number of version issues detected by \tool{}, PyEGo~\cite{9793962}, and Watchman~\cite{10.1145/3377811.3380426} on different DL stack layers}
    \label{fig: distribution}
\end{figure}

Table~\ref{table:issue_detection} shows the precison, recall, and F1 score of {\tool}, Watchman, and PyEGo. Overall, \tool{} achieves 91.7\% precision and 64.7\% recall, significantly outperforming Watchman and PyEgo. The success of \tool{} can be attributed to two factors. First, {\tool} is capable of detecting version issues across all five layers in a DL stack, while PyEGo and Watchman can detect version issues in at most two layers (i.e., library and runtime).  For example, an issue is caused by the incompatibility \texttt{Tensorflow 1.7.3} and {Cuda 11.2}, which involved the library layer and the driver layer. However, both Watchman and PyEGo do not contain version knowledge related to hardware drivers in their knowledge base, so they ignore this issue or report the latest library version that is compatible with other third-party libraries and the Python runtime version.  
In our benchmark, a majority of version issues involve the driver, OS/Container, and hardware layers. Figure~\ref{fig: distribution} shows the number of version issues detected by {\tool}, Watchman, and PyEgo on different DL stack layers. Second, \tool{} relies on version compatibility knowledge extracted from thousands of SO posts, which is more comprehensive and up-to-date than the information sources of PyEGo and Watchman.

\begin{table}[!h]
\centering
\caption{Accuracy of version incompatibility detection}
\label{table:issue_detection}
\begin{tblr}{
    colspec={lrrr},
    row{1} = {font=\bfseries},
    row{2-Z} = {rowsep=.3pt},
}
\toprule
\SetCell[c=1]{c} & \SetCell[c=1]{c}{\textsf{Precision}} & \SetCell[c=1]{c}{\textsf{Recall}} & \SetCell[c=1]{c}{\textsf{F1 score}} \\
\midrule
Watchman~\cite{10.1145/3377811.3380426} & 16.7\% & 5.9\% & 8.7\% \\
PyEGo~\cite{9793962} & 33.3\% & 29.4\% & 31.2\% \\
\tool{} & 91.7\% & 64.7\% & 75.9\% \\
\bottomrule
\end{tblr}
\end{table}

\subsection{Knowledge Graph Quality}
\label{sec: KG quality}
To evaluate the accuracy of the extracted knowledge in the knowledge graph produced by \tool{}, we randomly sampled 343 version (in)compatibility relations from 3,124 relations in the knowledge graph. This sample size is statistically meaningful with a 95\% confidence interval. For each relation, the first two authors independently verified whether the relation is true by searching online. Then, they compared their verification results and resolved any disagreement. The Cohen's Kappa score was 0.89. Overall,  287 of the 343 relations are verified to be correct, indicating a high accuracy (83.7\%). There are three reasons for the inaccuracy. First, 64\% of the incorrect relations were due to the mismatch between a component name and a version number. 27\% were due to incorrect predictions from UnifiedQA. 9\% were due to incorrect version knowledge shared in the original SO post.

\subsection{Pre-trained QA Model Performance}
\label{sec: qa performance}
When constructing the knowledge graph, UnifiedQA received a total of 5,532 queries (i.e., a question and a paragraph from SO) to predict compatibility relations between two DL components. Thus, to evaluate the accuracy of UnifiedQA, we randomly sampled 360 of the 5,532 queries. This sample size is statistically significant with a 95\% confidence interval. 
The first two authors independently validated each query result and discussed any disagreements. The Cohen's Kappa score for this step was 0.86. We found that UnifiedQA achieved 84.2\% precision and 91.3\% recall on these 360 queries. There are three reasons for incorrect predictions. First, 12\% of the incorrect predictions were because UnifiedQA misunderstood the question. Second, for 53\% of the incorrect predictions, the two DL components were just mentioned in the given paragraph and did not have dependency relationships. Thus, there is no version knowledge to extract, but UnifiedQA still predicts a relation. Finally, 35\% of the incorrect predictions were due to a mismatch of components and their versions in the query. This is more of an error propagated from the previous DL component recognition step. 

\subsection{Question Template Design}
\label{sec: QTanalysis}

Question prompt design is a classic problem when applying pre-trained QA models to downstream tasks~\cite{radford2021learning}. In this experiment, we designed 8 question templates, as shown in Table~\ref{table:question_template}. We measured the precision and recall of each template on the same sample of 360 relations as in Section~\ref{sec: qa performance}. Table~\ref{table:question templates results (single)} shows the model accuracy when using different question templates to extract version compatibility knowledge. Among eight question templates, Q7 (i.e., \textit{``Does A work with B?''}) achieves the best performance. 

\begin{table}[!h]
\caption{QA model accuracy of different question templates}
\label{table:question templates results (single)}
\resizebox{0.92\linewidth}{!}{
\begin{tabular}{lrrlrr}
\hline
\multicolumn{1}{c}{Question} & \multicolumn{1}{c}{Precision} & \multicolumn{1}{c}{Recall} & \multicolumn{1}{c}{Question} & \multicolumn{1}{c}{Precision} & \multicolumn{1}{c}{Recall} \\ \hline
Q1                             & 82.2\%                        & 89.2\%                      & Q5                            & 79.2\%                        & 85.8\%                      \\
Q2                             & 69.7\%                        & 75.6\%                      & Q6                            & 54.4\%                        & 59.0\%                      \\
Q3                             & 82.2\%                        & 89.2\%                      & \textbf{Q7}                   & \textbf{83.6\%}               & \textbf{90.7\%}             \\
Q4                             & 72.8\%                        & 78.9\%                      & Q8                            & 60.6\%                        & 65.7\%                      \\\hline
\end{tabular}}
\end{table}

Previous work has attempted to ask alternative questions to a QA model and ensemble the answers to improve the robustness and consistency of the QA model~\cite{radford2021learning}. Inspired by this, we also experimented with three different combination strategies to check if they indeed improve the model accuracy in our approach. The first combination strategy is to combine two questions with opposite phrases (e.g., \textit{``Is A compatible with B?'' and ``Is A not compatible with B?''}). The second combination strategy is to combine all question templates with positive phrases (Q1+Q3+Q5+Q7) and all templates with negative phrases (Q2+Q4+Q6+Q8). The third combination strategy is to simply combine all eight question templates. Given a combination of question templates, \tool{} will generate a corresponding set of questions and retrieve multiple answers from the UnifiedQA. 
For each question template, the UnifiedQA forwards the question together with the context and gets a cross-entropy loss value that measures how confident the model is about the answer. 
Then, it selects the one with the lowest loss value as the final answer, following the ensemble method in the previous work~\cite{radford2021learning}. Table~\ref{table: question templates results (combine)} shows the model accuracy when using different combinations of question templates. We found that combining multiple questions does not always lead to better performance compared with using a single question. For example, when combining questions with all positive phrases together (Q1+Q3+Q5+Q7), the precision and recall become worse compared with using Q7 alone. Among all combinations, combining Q1 and Q2 achieves the best performance (84.2\% precision and 91.3\% recall), while the improvement over the best individual template (Q7) is marginal.

\begin{table}[!h]
\caption{Accuracy of combined question templates}
\label{table: question templates results (combine)}
\resizebox{0.92\linewidth}{!}{
\begin{tabular}{lrrlrr}
\hline
\multicolumn{1}{c}{Question} & \multicolumn{1}{c}{Precision} & \multicolumn{1}{c}{Recall} & \multicolumn{1}{c}{Question} & \multicolumn{1}{c}{Precision} & \multicolumn{1}{c}{Recall} \\ \hline
\textbf{Q1 + Q2}               & \textbf{84.2\%}               & \textbf{91.3\%}             & Q5 + Q6                          & 70.8\%                        & 76.8\%                       \\
Q3 + Q4                        & 82.2\%                        & 89.2\%                      & Q7 + Q8                          & 82.8\%                        & 89.8\%                       \\ \hline 
Q1, 3, 5, 7                 & 82.5\%                        & 89.5\%                      & Q2, 4, 6, 8                   & 69.7\%                        & 75.6\%                       \\ \hline
Q1-8                 & 82.2\%                        & 89.2\%                     &                 &                        &                      \\
\hline
\end{tabular}
}
\end{table}

\subsection{Knowledge Consolidation Design}
The knowledge consolidation module of \tool{} aims to eliminate redundancies and reconcile conflicts in the prediction results of the QA model. We tried three different knowledge consolidation strategies:
(1) majority vote (i.e., select the relation predicted by the majority of posts); (2) weighted majority vote (i.e., select the relation predicted from the SO posts whose total SO score is the highest); (3) vote by loss  (i.e., select the predicted relation with the lowest loss value). In the resulting knowledge graph, 558 (in)compatibility relations were consolidated from multiple predicted relations. Thus, we randomly sampled 228 of the 558 relations and retrieved the original predictions from UnifiedQA. The sample size is statistically significant at a confidence interval of 95\%. Then, we evaluated the three strategies on this sample and compared the consolidation results to the ground truths, which were manually validated by the first author. Note that to fully evaluate the knowledge consolidation strategies, we excluded 25 incorrect relations that were caused by the mismatch between component names and versions. The results showed that the majority vote strategy achieves the best accuracy,  95.1\%, followed by voting by loss (93.6\%) and weighted majority vote (91.6\%). Yet the accuracy variance among these three strategies is small. This result implies that, unlike question template design, knowledge consolidation does not have a significant influence over the knowledge extraction pipeline. 

\section{Discussion}
\label{sec: discussion}
Our experiment results demonstrate the effectiveness of leveraging the rich version knowledge shared on Stack Overflow to detect version incompatibility issues in deep learning. Furthermore, we demonstrate the feasibility of leveraging pre-trained QA models to extract version compatibility knowledge from online discussions. This is significant since given the superior performance of pre-trained large models on text data, our approach can more accurately reason about the deep semantics in natural language narratives compared with rule-based systems. Our experiment results also demonstrate that, with careful prompt design, using the pre-trained QA model without finetuning can already reach a reasonable accuracy---84.2\% precision and 91.3\% recall as shown in Section~\ref{sec: qa performance}. 

In this work, we proposed a pipeline for generating a large-scale and high-quality knowledge graph, which is easily extensible with more SO posts. For now, \tool{} only builds the knowledge graph from SO posts concerning DL projects' version issues. Future researchers can extend this knowledge graph and apply it to other tasks by adding more diverse SO posts, such as detecting version issues in Java projects by extending the knowledge graph with SO posts concerning Java ecosystem.

\noindent\textbf{\emph{Comparison to ChatGPT.}} Given the recent advances in Large Language Models (LLMs) such as ChatGPT, one alternative solution is to directly ask LLMs whether two versioned libraries are compatible or not. We investigated this by prompting ChatGPT (GPT 3.5) with the template \texttt{``is [Lib A] compatible [Lib B]?''}. Specifically, we tested ChatGPT on all 17  pairs of incompatible DL components from our benchmark, as well as 290 randomly sampled 
 pairs of compatible components. We found out that ChatGPT correctly predicted 12 incompatible relations and 211 compatible relations. While ChatGPT achieves a decent recall (70.5\%) on the benchmark, it achieves a much lower precision (72.6\%) compared to our tool (91.7\%). This finding suggests that ChatGPT indeed obtains some DL compatibility knowledge. However, ChatGPT may still generate hallucinated knowledge, which prevents it from accurately identifying version issues in DL projects.


\noindent\textbf{\emph{Threats to validity.}} 
In terms of internal validity, one threat is the small benchmark used to evaluate \tool{}. So far, this benchmark contains only 10 DL projects from GitHub. While these projects are popular projects (at least 100 stars) from 3 different DL domains, our experiment can still benefit from a larger benchmark with more domains. 
In the future, the authors will expand the current benchmark to include more DL projects. Furthermore, given the large number of processed SO posts, we cannot manually validate each one of them to evaluate the accuracy of our knowledge extraction pipeline. Thus, we inspected random samples, which may lead to imprecise estimation. We mitigate this threat by using a statistically significant sample size at the 95\% confidence interval.

In terms of external validity, the current implementation of \tool{} only supports 48 popular DL stack components. While \tool{} has been demonstrated to effectively extract version knowledge of these popular components, it may not be able to extract useful knowledge for DL components that are rarely discussed on Stack Overflow, which diminishes its effectiveness for version compatibility detection for those components.

In terms of construct validity, we cannot guarantee the DL stack component recognition algorithm always correctly assigns the version numbers to the intended component. Consider the example: \textit{``Answer v2.0: Try to install tensorflow 1.15 on your Python.''} Since the current implementation of \tool{} is designed to match as many component-version pairs as possible, \texttt{v2.0} will be incorrectly matched with \texttt{Python}, although it only indicates we are reading the second edition of this answer.

\noindent\textbf{\emph{Future work.}} There are several interesting future directions to explore for this problem. First, \tool{} currently only extracts knowledge from SO posts. Other types of documents, such as PyPI documentation and GitHub issue pages, also contain knowledge relevant to library versions. 
Thus, it is worthwhile extending \tool{} to extract knowledge from other types of online documents. Specifically, the knowledge consolidation process needs to be redesigned to merge knowledge extracted from different types of documents while accounting for their credibility, recency, and popularity. 

Second, \tool{} can only detect version incompatibilities without automatically repairing the version issues in DL projects. However, the knowledge graph generated from SO discussions contains knowledge of compatible relations between DL components. In the future, we plan to investigate effective repair strategies based on the knowledge graph. 

Finally, although the pre-trained QA model in {\tool} can extract version knowledge with reasonable accuracy (84.2\% precision and 91.3\% recall), its accuracy could be further improved with fine-tuning. In the future, we plan to fine-tune UnifiedQA by creating a large set of SO posts labeled version compatibility relationships. Alternatively, we can also substitute UnifiedQA with stronger LLMs, such as ChatGPT, to extract version knowledge from SO posts. Recent work  has demonstrated that ChatGPT achieves good performance in knowledge extraction~\cite{wei2023zero, li2023evaluating}. Therefore, we believe it is a promising direction to improve \tool{} performance by substituting UnifiedQA with ChatGPT.



\section{Related Work}

\subsection{Version Incompatibility Detection}
There is a large body of literature on version incompatibility detection~\cite{9000001, 8812128, 10.1145/3180155.3180184, wang2018dependency, 10.1145/3377811.3380426, horton2019dockerizeme, mukherjee2021fixing, wang2021restoring, 9793962}. Those most related to our work are techniques designed for the Python ecosystem~\cite{10.1145/3377811.3380426, 9793962, wang2021restoring}. Wang et al.~\cite{10.1145/3377811.3380426} proposed WatchMan, which collects the metadata of each PyPI project and constructs a knowledge base. It generates and traverses a dependency graph to look for version incompatibilities among a given list of third-party Python packages. 
Ye et al.~\cite{9793962} proposed PyEGo, which automatically infers the latest compatible versions for required dependencies in Python projects. They constructed a knowledge graph that stores relations and constraints among third-party Python packages, the Python interpreter, and system libraries. The knowledge is collected from official documents such as PyPI and Python official website. SnifferDog~\cite{wang2021restoring} constructs an API bank that maps APIs to different versions of Python packages. With this bank, SnifferDog infers required package versions for Jupyter notebooks from API usage.
Since these techniques only extract dependency relations from official documents such as PyPI metadata, they cannot detect errors in the driver, OS/container, and hardware layer in a DL stack. By contrast, \tool{} uses a pre-trained QA model to extract version knowledge from online discussions, which are more comprehensive and up-to-date. 

{There are also some version incompatibility detection approaches for other ecosystems such as Java and JavaScript~\cite{10.1145/3180155.3180184, article12018, 9000001}. 
Patra et al.~\cite{10.1145/3180155.3180184} proposed ConflictJS, which detects conflicts in JavaScript libraries by identifying libraries that write to the same global memory location. He et al.~\cite{9000001} proposed IctApiFinder, which uses an integer-procedural data flow analysis framework to identify incompatible API usages in Android applications. However, these techniques cannot be applied to Python-based DL projects due to the language difference.}
\subsection{Knowledge Extraction from SE Documents}
Several approaches have been proposed to extract API knowledge or insightful sentences from API documents and SO discussions~\cite{liu2019generating, li2018improving, wang2019extracting, treude2016augmenting, yin2021api, liu2021api, xu2016predicting}.
Some of them perform knowledge extraction with rule-based pattern matching~\cite{li2018improving, liu2021api}. For example, 
Li et al.~\cite{li2018improving} developed a set of linguistic patterns to extract ten types of API usage sentence-level caveats from SO posts. Recently, more techniques have sought to 
improve their flexibility in processing SE documents via neural networks. 
For example, Liu et al.~\cite{liu2019generating} trained a feed-forward neural network to classify descriptive sentences of APIs from API documentation.
Similarly, DeepTip~\cite{wang2019extracting} uses a CNN model to extract sentence-level API usage tips from SO posts with a trained CNN model.
Compared with previous work, our approach performs a more fine-grained knowledge extraction that requires a delicate recognition of DL components followed by inference of their relationship, rather than classifying sentences. Furthermore, previous neural approaches need to acquire a large labeled dataset first and train a model from the scratch. However, our approach makes use of a pre-trained QA model and demonstrates the feasibility of achieving a reasonable accuracy without finetuning through careful question template design.

\section{Conclusion}
This paper presents \tool{}, a knowledge-based version incompatibility detection approach for deep learning projects. The key insight is to leverage the abundant version compatibility knowledge from Stack Overflow to facilitate the detection of version incompatibilities. Specifically, {\tool} uses a pre-trained Question-Answering (QA) model to extract version compatibility knowledge from the free-form text in online discussions. Compared with existing rule-based knowledge extraction systems, utilizing a pre-trained QA model empowers {\tool} to reason about the deep semantics in natural language without the need of acquiring domain-specific data to train a model from scratch. The evaluation results demonstrate that our approach can extract version knowledge with 84\% accuracy and can accurately identify 65\% of known version issues in 10 popular DL projects with a 92\% precision, significantly outperforming two state-of-the-art approaches.


\section*{Acknowledgment}
The authors would like to thank the anonymous reviewers for their valuable comments. This research was in part supported by an Amazon Research Award and a Cisco Research Award.


\bibliographystyle{ACM-Reference-Format}
\bibliography{reference}


\end{document}